\documentclass[11pt]{article}

\usepackage[margin=1in]{geometry}
\usepackage{graphicx}
\usepackage{amsmath,amssymb}
\usepackage{booktabs}
\usepackage{authblk}
\usepackage{hyperref}
\usepackage{caption}
\usepackage{subcaption}
\usepackage{tabularx}
\usepackage{booktabs}
\usepackage{siunitx}
\usepackage[utf8]{inputenc}   % for old TeXLive; harmless otherwise
\usepackage[T1]{fontenc}      % enables \k and many accented chars

\title{Why Nonlinear Models Matter: Unified Analysis of Cognitive Load, Stress, and Exercise Using Wearable Physiological Signals}

\author[1]{Khondakar Ashik Shahriar\thanks{Corresponding author: \texttt{kh.ashikshahriar@gmail.com}}}
\affil[1]{Department of Electrical and Electronic Engineering, Bangladesh University of Engineering and Technology (BUET), Dhaka, Bangladesh}

\date{} % No date

\begin{document}

\maketitle

\begin{abstract}
Wearable physiological signals exhibit strong nonlinear and subject-dependent behavior, challenging traditional linear models. This study provides a unified evaluation of cognitive load, stress, and physical exercise recognition using three public Empatica~E4 datasets. Across all conditions, nonlinear machine learning models consistently outperformed linear baselines, achieving 0.89--0.98 accuracy and 0.96--0.99 ROC--AUC, while linear models remained below 0.70--0.73 AUC. Although Leave-One-Subject-Out validation revealed substantial inter-individual variability, nonlinear models maintained moderate cross-person generalization. Ablation and statistical analyses confirmed the necessity of multimodal fusion, particularly EDA, temperature, and ACC, while SHAP interpretability validated these findings by uncovering physiologically meaningful feature contributions across tasks. Overall, the results demonstrate that physiological state recognition is fundamentally nonlinear and establish a unified benchmark to guide the development of more robust wearable health-monitoring systems.
\end{abstract}

\section{Introduction}

Wearable devices have rapidly evolved from simple step counters into sophisticated physiological monitors capable of capturing blood volume pulse (BVP), electrodermal activity (EDA), skin temperature, motion dynamics, and related autonomic indicators~\cite{jegan2024development,wu2018materials}. As these sensors become increasingly embedded in daily life, they offer an unprecedented opportunity to infer cognitive and physiological states unobtrusively. Detecting when an individual is stressed, cognitively overloaded, or engaging in varying levels of physical effort is essential for next-generation applications in digital health~\cite{canali2022challenges,stern2022advancing}, human--computer interaction~\cite{chen2017wearable}, and personalized well-being interventions~\cite{yen2021smart}. Yet the promise of wearables remains constrained by a fundamental challenge: physiological responses are nonlinear~\cite{baxt1994complexity}, context-dependent~\cite{khalaf2020analysis}, and vary widely across individuals~\cite{chen2017subject,can2023approaches}.

Traditional linear classification models assume stable and linearly separable relationships between input signals and target states~\cite{hastie2008linear}. However, the human autonomic nervous system does not operate linearly~\cite{kowalik1996does}. Stress induces rapid sympathetic activation reflected in phasic electrodermal activity bursts; cognitive load modulates heart-rate variability in subtle, subject-specific ways; and aerobic versus anaerobic exercise produces complex shifts in BVP-derived cardiovascular features and motion dynamics. These interactions create curved, overlapping, and non-stationary decision boundaries that linear models struggle to capture, leading to inconsistent results across subjects and datasets.

Recent advances in the availability of large-scale, publicly accessible wearable-sensor datasets have created new opportunities to study physiological dynamics across a wide range of human states. Collections capturing mental workload, emotional arousal, stress reactivity, and physical activity provide diverse yet complementary windows into autonomic regulation~\cite{nandini2025ensemble,liu2025signparser,sztyler2017position,piciucco2021biometric}. Although these datasets originate from different experimental contexts, they are typically built upon a shared set of multimodal signals, most commonly photoplethysmography (PPG), EDA, peripheral temperature, and multi-axis accelerometry. The recurrence of these modalities across independent datasets enables a rare, cross-domain perspective: it becomes possible to evaluate how various computational models respond to the underlying nonlinearities inherent in physiological time series, rather than to the idiosyncrasies of any single task or cohort. This convergence positions wearable-sensor research at a pivotal moment, where unified methodological investigations can uncover general principles of physiological modeling and inform the design of more robust, generalizable systems.

This study examines whether nonlinear machine learning models offer a significant advantage over linear baselines in detecting cognitive load, stress, and exercise states from wearable signals. By applying consistent preprocessing, feature extraction, model evaluation, Leave-One-Subject-Out (LOSO) validation~\cite{elisseeff2003leave}, and SHAP-based interpretability~\cite{lundberg2017unifiedapproachinterpretingmodel} across three independent datasets, the analysis aims to uncover patterns that persist beyond individual experimental settings. The findings demonstrate that physiological state recognition is inherently nonlinear, and that nonlinear learners capture meaningful physiological markers, such as EDA reactivity, BVP-derived cardiovascular shifts, and temperature--motion synergies, far more effectively than linear approaches. While deep learning models are increasingly popular for physiological state recognition, they require high computational resources and a large subject cohort for generalization, making them impractical for real-time inference on wearable devices. Therefore, this work focuses on benchmarking nonlinear classical machine-learning models, which provide promising accuracy while remaining lightweight and suitable for on-device deployment.

The contributions of this work are summarized as follows:
\begin{itemize}
    \item It introduces a unified nonlinear learning framework to jointly analyze cognitive load, stress, and exercise using multimodal wearable physiological signals.
    \item It demonstrates that nonlinear machine learning models provide substantially improved recognition performance with corresponding latent space visualization for complex, dynamic physiological states compared to traditional linear approaches.
    \item It proposes a SHAP-based interpretability strategy that reveals physiologically meaningful biomarkers and offers insight into how each signal contributes to inference.
    \item It establishes strong subject-independent generalization through LOSO validation across participants and robustness of performance through rigorous ablation and associated statistical analysis.
\end{itemize}

\section{Related Work}

Wearable physiological sensing enables unobtrusive monitoring of cognitive, emotional, and physical states using signals such as BVP, EDA, temperature, and accelerometry, offering rich insights into autonomic activity. However, reliably mapping these signals to human states remains difficult due to their nonlinear, highly individualized dynamics~\cite{roos2023wearable,seneviratne2017survey}. Autonomic responses exhibit complex interactions across modalities, such as phasic EDA bursts~\cite{posada2019phasic}, subject-specific HRV patterns, and coupled cardiovascular--motion effects~\cite{schulz2013cardiovascular}, that violate assumptions of linearity and stationarity. Consequently, linear models like logistic regression, LDA, and linear SVMs often fail to generalize across subjects or tasks. These limitations have led to growing adoption of nonlinear approaches that better capture higher-order dependencies in physiological signals.

Several datasets highlight the strengths of nonlinear learners and the limitations of linear approaches. A recent \textit{Scientific Data} publication introduced a large, rigorously collected dataset containing acute stress induction sessions, as well as aerobic and anaerobic exercise, all recorded noninvasively with the Empatica E4~\cite{he0v-tf17_dataset,hongn2025wearable}. Using data from 36 stress, 30 aerobic, and 31 anaerobic participants, the authors showed that nonlinear models like XGBoost achieved high accuracy (93\% stress--rest, 91\% aerobic--anaerobic, 84\% four-class), highlighting the complex, nonlinear nature of physiological state boundaries. Complementary evidence comes from real-world stress research, where the goal is often to capture naturalistic stress patterns rather than laboratory-induced responses. One such study presented a publicly available dataset collected during academic exams, in which skin conductance and skin temperature signals were recorded from 10 subjects across three exam sessions~\cite{Amin2022_WearableExamStress,9744065}. They showed that coarse-grained EDA trendline features enabled 70--80\% accuracy in predicting performance, demonstrating both the feasibility and challenges of wearable stress monitoring in naturalistic settings. Lazarou et al.~\cite{article} provide a concise review of real-time stress prediction from wearable physiological data, cataloguing sensors, ML methods, commercial products, and open datasets while highlighting challenges (ground truth, motion/noise, personalization, device heterogeneity) and setting a research agenda for robust, real-world stress monitoring. Several studies~\cite{pandukabhaya2025performance,papi2015use,liu2018benchmark,xu2020stochastic,zhang2022can} have also benchmarked stress-related datasets and demonstrated how machine-learning models, particularly nonlinear approaches, substantially improve physiological stress recognition and advance the development of reliable wearable monitoring systems.

Current research remains fragmented, with stress, cognitive load, and exercise studied separately using inconsistent pipelines and limited cross-dataset or subject-independent evaluation. Few works~\cite{ramadan2024multimodal,khandelwal2025machine} systematically compare linear and nonlinear models or analyze cross-domain physiological mechanisms through unified interpretability tools. Motivated by these gaps, the present study provides a unified, cross-dataset assessment of linear vs.\ nonlinear modeling, demonstrating the fundamentally nonlinear nature of physiological state recognition and identifying robust markers that generalize across domains.

\section{Methods}

\subsection{Theoretical Motivation}

Physiological signals acquired from wearable devices arise from coupled nonlinear dynamical processes driven by the interaction between the sympathetic and parasympathetic branches of the autonomic nervous system (ANS). These dynamics violate the linearity assumptions underlying classical statistical learning models, thereby motivating the use of nonlinear classification functions. The following subsections summarize the core theoretical principles supporting this choice.

\subsubsection*{1) Nonlinear Autonomic Dynamics}

Let $x_{\mathrm{HR}}(t)$, $x_{\mathrm{EDA}}(t)$, $x_{\mathrm{TEMP}}(t)$, and $x_{\mathrm{ACC}}(t)$ denote the recorded physiological modalities. Their temporal evolution is governed by nonlinear differential equations of the form
\begin{equation}
\frac{dx(t)}{dt} = f\big(x(t), u(t)\big),
\end{equation}
where $u(t)$ represents latent sympathetic and parasympathetic inputs. For instance, heart-rate variability (HRV) is shaped by the baroreflex control loop~\cite{watkins1995assessment}, which is frequently modeled as a second-order nonlinear oscillator~\cite{goldberger2002fractal}:
\begin{equation}
\frac{d^2 x_{\mathrm{HR}}}{dt^2}
+ \alpha(x_{\mathrm{HR}})\frac{dx_{\mathrm{HR}}}{dt}
+ \beta(x_{\mathrm{HR}}) = \gamma\, x_{\mathrm{BP}}(t),
\end{equation}
with state-dependent nonlinear coefficients $\alpha(\cdot)$ and $\beta(\cdot)$. Electrodermal activity (EDA) exhibits threshold-based phasic responses that can be expressed as
\begin{equation}
x_{\mathrm{EDA}}(t)=
\begin{cases}
x_{\mathrm{EDA}}(t^-) + A e^{-\lambda (t-t_0)}, & \text{if } u(t_0)>\theta,\\[4pt]
x_{\mathrm{EDA}}(t^-), & \text{otherwise},
\end{cases}
\end{equation}
producing piecewise-nonlinear and discontinuous dynamics~\cite{benedek2010decomposition}. These intrinsic nonlinearities imply that the mapping from underlying physiological state to observable signal is non-affine and therefore cannot be well approximated by linear decision boundaries.

\subsubsection*{2) Nonlinear Feature Interactions}

Let $\mathbf{z}\in\mathbb{R}^d$ denote the handcrafted feature vector extracted from a window of multimodal data. For stress ($S$), cognitive load ($C$), or exercise ($E$), the psychophysiological response can be expressed as
\begin{equation}
y = g(\mathbf{z}) + \varepsilon,
\end{equation}
where $g(\cdot)$ is a nonlinear function capturing the underlying ANS dynamics. Empirical observations indicate that $g$ includes multiplicative interactions between features, such as
\begin{equation}
g(\mathbf{z}) =
w_1 z_{\mathrm{EDA}} + 
w_2 z_{\mathrm{HR}} +
w_3 z_{\mathrm{TEMP}} +
w_4 z_{\mathrm{ACC}} +
w_5 z_{\mathrm{EDA}}z_{\mathrm{HR}} +
w_6 z_{\mathrm{ACC}}^{2} + \cdots,
\end{equation}
implying that $g$ is at least quadratic. More generally, ANS-driven processes admit Volterra-series expansions~\cite{marmarelis2004nonlinear}:
\begin{equation}
g(\mathbf{z})=
\sum_{i} a_i z_i
+ \sum_{i,j} a_{ij} z_i z_j
+ \sum_{i,j,k} a_{ijk} z_i z_j z_k
+ \cdots,
\end{equation}
which cannot be represented by linear classifiers that assume an affine mapping.

\subsubsection*{3) Why Linear Models Fail: Margin Geometry}

A linear classifier $h(\mathbf{z})=\mathbf{w}^\top\mathbf{z}+b$ requires linearly separable feature sets, i.e.,
\begin{equation}
y_i(\mathbf{w}^\top\mathbf{z}_i+b) > 0
\quad \forall i.
\end{equation}
However, low-dimensional embeddings of the extracted features (e.g., t-SNE or UMAP) reveal non-convex, intertwined manifolds for the target classes. Formally,
\begin{equation}
\mathrm{conv}(\mathcal{Z}_S) \cap \mathrm{conv}(\mathcal{Z}_C) \neq \varnothing,
\end{equation}
where $\mathcal{Z}_y$ denotes the feature set for class $y$. When convex hulls overlap, no linear classifier can achieve perfect separation, regardless of regularization or feature scaling.

\subsubsection*{4) Nonlinear Kernels Approximate the True Physiological Function}

Nonlinear classifiers such as RBF-SVMs and multilayer perceptrons implicitly perform feature lifting into high-dimensional spaces~\cite{ring2016approximation}:
\begin{equation}
\phi:\mathbb{R}^{d}\rightarrow\mathcal{H}, \qquad
K(\mathbf{z}_i,\mathbf{z}_j)=\langle\phi(\mathbf{z}_i),\phi(\mathbf{z}_j)\rangle,
\end{equation}
where $\mathcal{H}$ is a Hilbert space. The RBF kernel
\begin{equation}
K(\mathbf{z}_i,\mathbf{z}_j)=
\exp\!\left(-\frac{\|\mathbf{z}_i-\mathbf{z}_j\|^{2}}{2\sigma^{2}}\right),
\end{equation}
acts as a universal approximator for continuous nonlinear functions and is therefore well-suited to model mappings of the form
\begin{equation}
y \approx g(\mathbf{z}),
\end{equation}
where $g$ contains high-order ANS interactions.

Given the nonlinear dynamics of autonomic physiology, the multiplicative structure of multimodal feature interactions, and the non-convex geometry of the resulting feature manifolds, linear models are theoretically inadequate for distinguishing stress, cognitive load, and exercise states. Nonlinear methods, such as RBF-SVMs, multilayer perceptrons, Random Forests, and gradient-boosted decision trees, are therefore more appropriate for capturing the true psychophysiological mapping and provide a principled foundation for the modeling choices adopted in this study.

\subsection{Datasets}

Three publicly available wearable physiology datasets were used to analyze cognitive load, psychological stress, and physical exercise states. Although collected under different experimental protocols, all datasets include core Empatica~E4 signals: EDA, BVP, HR, skin temperature, and tri-axial accelerometry (ACC), providing a consistent feature space for unified cross-dataset analysis.

\subsubsection{Wearable Stress and Exercise Dataset}

The dataset~\cite{he0v-tf17_dataset} contains multimodal physiological signals (EDA, BVP, HR, skin temperature, and 3-axis acceleration) recorded with the Empatica E4 during structured stress induction, aerobic, and anaerobic exercise protocols. Data were collected from healthy adults aged 18--30 across 36 stress sessions, 30 aerobic sessions, and 31 anaerobic sessions. The stress protocol included cognitive, emotional, and arithmetic stressors with rest periods and self-reported stress ratings, while the exercise protocols involved controlled cycling routines with varying intensity, including progressive aerobic workloads and high-intensity anaerobic sprints. All sessions include raw sensor data with event markers, demographic data, and self-reported stress measures.

\subsubsection{Exam Stress Dataset}

This dataset~\cite{Amin2022_WearableExamStress} provides multimodal physiological recordings collected from college students wearing the FDA-approved Empatica E4 wristband during three real-world academic exams (Midterm 1, Midterm 2, and Final). Unlike traditional laboratory stress datasets that use artificial stimuli, this dataset captures electrodermal activity, heart rate, blood volume pulse, temperature, inter-beat intervals, and accelerometer signals during authentic cognitive stressors lasting 1.5 to 3 hours. Each participant selected an E4 device before the exam, enabling anonymized linking of physiological recordings with exam grades. The dataset includes raw sensor CSV files, timestamps, metadata, and grade information for ten students, providing a unique opportunity to study real-world exam stress and its potential relationship with academic performance. To enable binary classification of academic performance, final exam scores were converted into performance labels. A performance threshold of 160 points was used to binarize final exam scores, labeling participants scoring $\geq 160$ as High performers (S1, S2, S3, S6, S8) and those scoring $<160$ as Low performers (S4, S5, S7, S9, S10), enabling direct mapping of physiological responses to academic outcomes. Despite its small sample size and the presence of motion artifacts inherent to naturalistic settings, the dataset serves as a valuable resource for developing stress and performance-prediction models using multimodal wearable data.

\subsubsection{Cognitive Load Dataset (CogWear)}

The CogWear dataset~\cite{grzeszczyk2023cogwear} comprises multimodal physiological recordings collected from consumer-grade and research-grade wearable devices, including the Empatica E4, Samsung Galaxy Watch4, and Muse S EEG headband, during controlled cognitive load experiments. It consists of two parts: a pilot dataset with 11 participants performing Stroop tasks and baseline rest, and a survey gamification experiment with 13 participants completing Stroop tests, resting phases, and a series of mobile surveys presented in gamified and non-gamified formats across two sessions. The dataset includes synchronized PPG, EDA, temperature, accelerometry, and EEG signals, along with detailed Stroop responses and survey interaction logs, enabling multimodal cognitive load estimation. Physiological signals were collected under standardized device placement and time-alignment procedures, and data are organized per participant into baseline, cognitive load, and survey segments. Only the Empatica-E4 data were used for the experiments.

\begin{table}[t]
\centering
\caption{Comparison of wearable datasets used in this study.}
\label{tab:dataset_comparison}
\begin{tabular}{lccc}
\hline
\textbf{Dataset} & \textbf{Classes} & \textbf{Duration} & \textbf{Modalities} \\
\hline
Stress--Exercise Dataset & Stress / Aerobic / Anaerobic &
$\sim$60\,min & EDA, TEMP, HR, IBI, ACC \\
Exam Performance Dataset & High vs Low Performance &
Final exam task & EDA, TEMP, HR, IBI, ACC, BVP \\
Cognitive Load Dataset & Baseline vs Load &
10--15\,min & EDA, TEMP, BVP, ACC \\
\hline
\end{tabular}
\end{table}

\subsection{Signal Visualization}

To qualitatively examine underlying physiological patterns and inter-subject variability, the mean and standard deviation of each modality were computed across subjects within each dataset. The visualizations highlight both condition-dependent responses and substantial between-subject variability.

\begin{figure}[!htbp]
    \centering
    \includegraphics[width=0.95\textwidth]{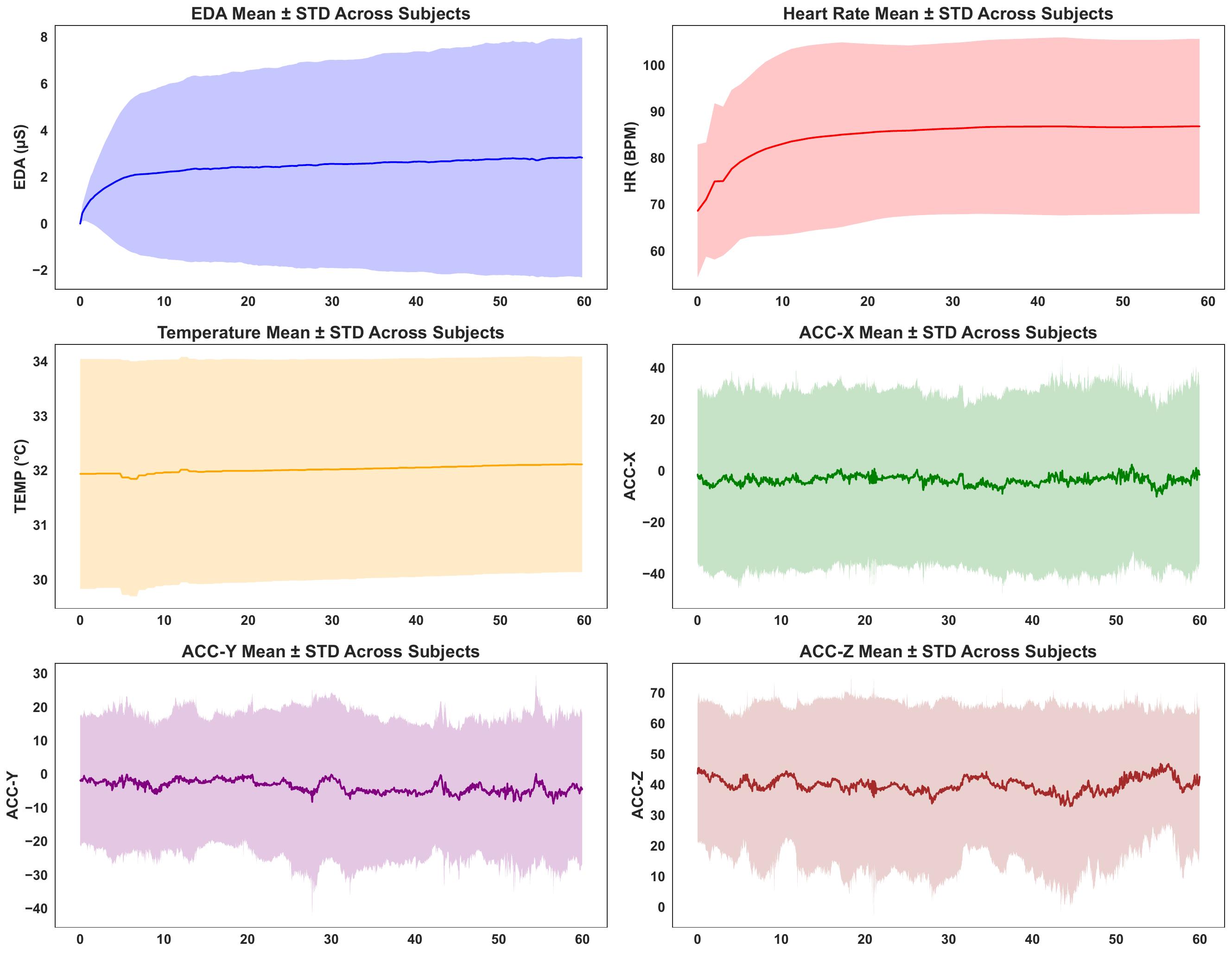}
    \caption{Mean $\pm$ standard deviation of EDA, HR, temperature, and accelerometer signals across subjects for Dataset~1. Substantial inter-subject variability is observable across all modalities.}
    \label{fig:dataviz1}
\end{figure}

\begin{figure}[!htbp]
    \centering
    \includegraphics[width=0.95\textwidth]{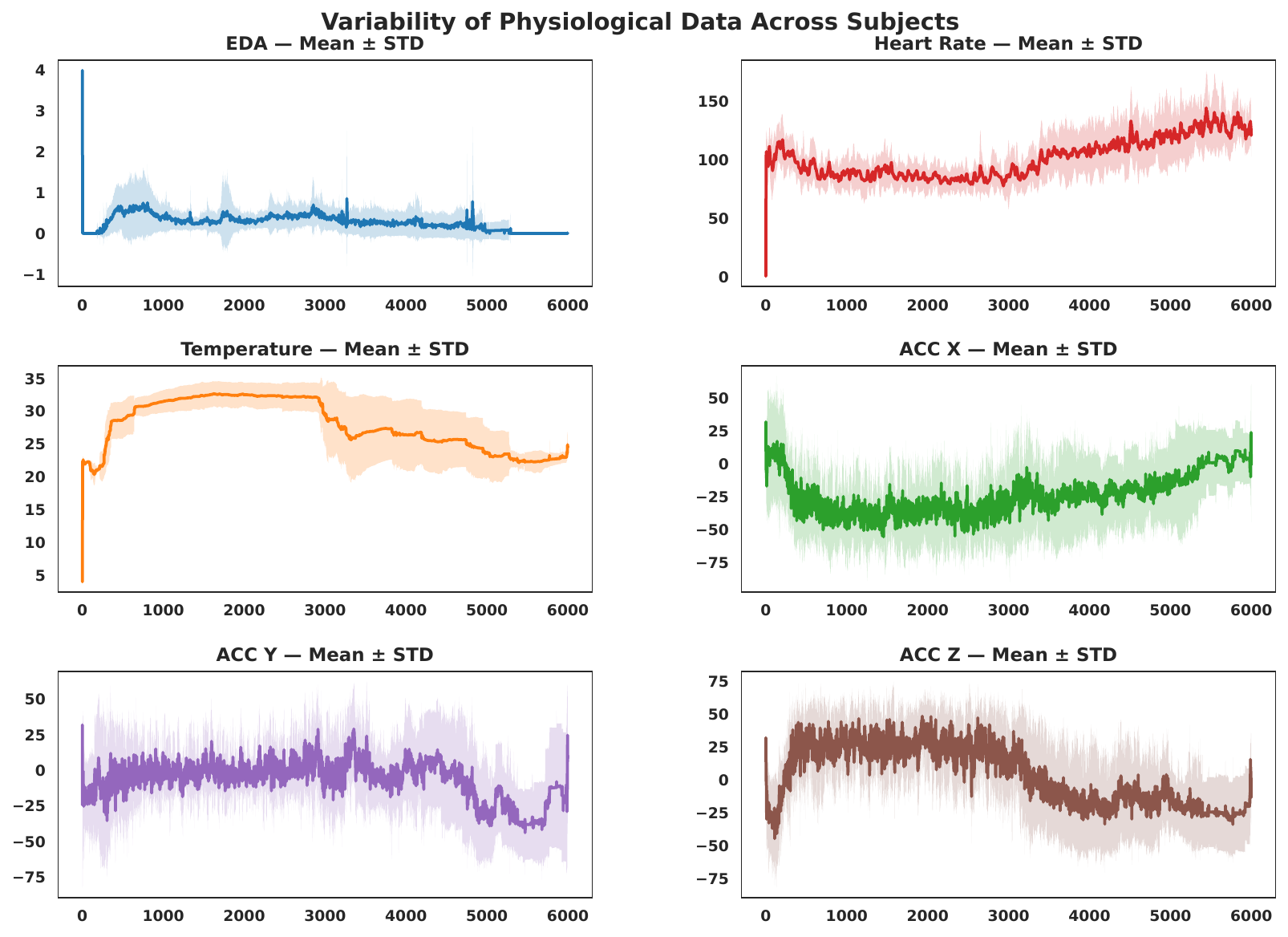}
    \caption{Physiological variability for Dataset~2. Aerobic and anaerobic exercise produce strong activation patterns in all modalities, particularly HR and ACC.}
    \label{fig:dataviz2}
\end{figure}

\begin{figure}[!htbp]
    \centering
    \includegraphics[width=0.95\textwidth]{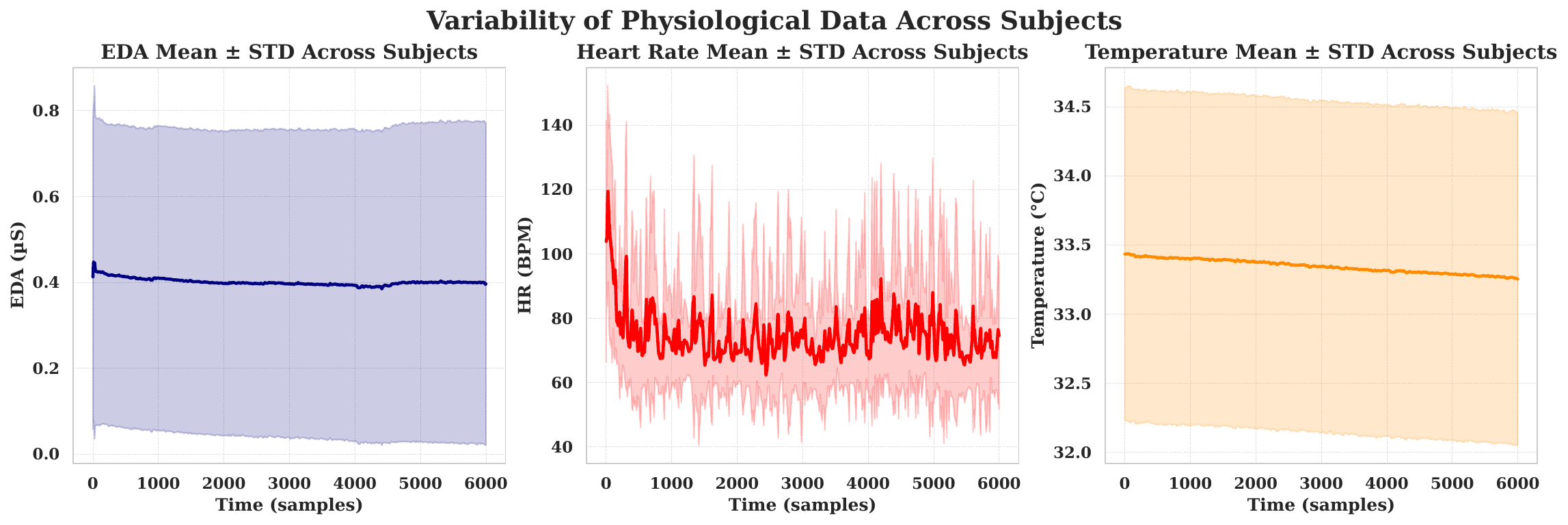}
    \caption{Cross-subject physiological variability for Dataset~3, illustrating subtle EDA and HR changes during academic exam stress and low-motion accelerometer behaviour.}
    \label{fig:dataviz3}
\end{figure}

\subsubsection*{Electrodermal Activity (EDA)}

EDA signals showed slow tonic drifts and intermittent phasic peaks across datasets. The shaded standard deviation regions in Figures~\ref{fig:dataviz1}--\ref{fig:dataviz3} indicate strong heterogeneity in sympathetic arousal across individuals.

\subsubsection*{Heart Rate (HR)}

HR exhibited rapid increases during exercise, moderate fluctuations during cognitive stress, and relatively stable trajectories during baseline periods. High variability confirms strong subject-specific cardiovascular dynamics.

\subsubsection*{Skin Temperature}

Temperature demonstrated smooth, gradual shifts influenced by both stress-induced vasoconstriction and exercise-induced thermogenesis. Though less variable than EDA or HR, it remained subject-dependent.

\subsubsection*{Accelerometry (ACC)}

Accelerometer patterns clearly separated sedentary cognitive tasks from physical activity. Aerobic segments produced periodic motion, anaerobic activity resulted in high-variance peaks, and cognitive tasks showed low-motion fluctuations. Across all modalities, wide standard deviation envelopes revealed considerable individual differences in physiological response intensity, baseline levels, and artefact behaviour. These patterns reinforce the need for nonlinear models and subject-independent evaluation strategies.

\subsection{Preprocessing and Feature Extraction}

Physiological signals from all three wearable datasets were processed using a unified pipeline to ensure consistency across stress, cognitive load, and exercise conditions. The Empatica~E4 device provides synchronized measurements including EDA, skin temperature, BVP, HR, IBI, and ACC. For subject $i$, the multivariate time series is defined as
\[
\mathbf{x}_i(t)=\big[x_i^{\mathrm{EDA}}(t),\,x_i^{\mathrm{TEMP}}(t),\,x_i^{\mathrm{HR}}(t),\,x_i^{\mathrm{IBI}}(t),\,\mathbf{x}_i^{\mathrm{ACC}}(t),\,x_i^{\mathrm{BVP}}(t)\big], 
\quad t=1,\dots, T_i.
\]
Labels differ depending on the dataset: (i) stress/aerobic/anaerobic activity, (ii) baseline vs.\ cognitive load, and (iii) high vs.\ low performance in an examination task.

\subsubsection*{Signal Cleaning, Synchronization, and Normalization}

Recordings were screened for missing files, empty IBI streams, and sensor dropouts. Each modality was standardized at the subject level to reduce inter-subject variability:
\[
\tilde{x}(t)=\frac{x(t)-\mu_x}{\sigma_x+\epsilon},
\]
where $\mu_x$ and $\sigma_x$ denote the subject-specific mean and standard deviation. Because sampling frequencies differ (4~Hz EDA/TEMP, 32~Hz ACC, 64~Hz BVP, 1~Hz HR), signals were aligned through time-based segmentation rather than equal-length resampling.

\subsubsection*{Temporal Segmentation}

Temporal segmentation plays a crucial role in transforming long-duration physiological recordings into structured samples that capture transient autonomic patterns relevant for stress and workload modeling~\cite{zhao2016decomposing}. Short, overlapping windows increase the number of training instances, enhance the model’s ability to detect rapid physiological changes~\cite{ferdousi2021windowed}, and ensure better synchronization across modalities with different sampling frequencies, while also limiting the effect of noise bursts or brief motion artifacts. In this work, continuous signals were segmented into overlapping windows so that, for subject $i$, modality $m$, and window index $k$, each segment is defined as
\[
\mathbf{x}^{(m)}_{i,k}=\{x_i^{(m)}(t)\mid t\in[kS,\,kS+W]\},
\]
with window length $W$ and stride $S$. Inter-beat interval (IBI) segments were obtained using timestamp masks due to the event-based nature of beat detections. Only windows containing valid samples across all modalities were retained. Datasets~1 and~2 used $W=30$\,s and $S=15$\,s, consistent with common practice in wearable-based stress research, whereas Dataset~3 employed shorter windows ($W=10$\,s, $S=5$\,s) to compensate for its smaller sample size and to preserve sufficient temporal variability for downstream classification.

\subsubsection*{Feature Engineering}

A unified set of statistical, cardiovascular, electrodermal, thermal, and motion-derived descriptors was extracted from each window. Table~\ref{tab:dataset_features} summarizes the feature sets and availability across datasets.

\begin{table}[t]
\centering
\caption{Feature sets extracted from each wearable modality.}
\label{tab:dataset_features}
\begin{tabular}{lccc}
\hline
\textbf{Modality} & \textbf{Features} & \textbf{Count} & \textbf{Used In} \\
\hline
EDA (4 Hz) & Mean, Std, Slope, SCR peaks & 4 & All datasets \\
TEMP (4 Hz) & Mean, Std & 2 & All datasets \\
HR / IBI (1 Hz) & Mean, Std, SDNN, RMSSD & 4 & Datasets 1 \& 2 \\
BVP (64 Hz) & Peak HRV, Amplitude, Energy & 2--4 & Datasets 2 \& 3 \\
ACC (32 Hz, 3 axes) & Mean, Std per axis & 6 & All datasets \\
\hline
\end{tabular}
\end{table}

\paragraph{Electrodermal Activity (EDA).}
Four features were computed:
\[
f_{\mathrm{EDA}}=[\mu,\,\sigma,\,\beta,\,P],
\]
where $\mu$ and $\sigma$ denote the mean and standard deviation, $\beta$ is the linear slope over 30\,s, and $P$ is the number of detected SCR peaks.

\paragraph{Skin Temperature (TEMP).}
Two features represent thermoregulatory variation:
\[
f_{\mathrm{TEMP}}=[\mu_{\mathrm{temp}},\,\sigma_{\mathrm{temp}}].
\]

\paragraph{Heart Rate and HRV from IBI (Datasets 1 \& 2).}
HR and IBI windows produce:
\[
f_{\mathrm{HRV}}=[\mu_{\mathrm{HR}},\,\sigma_{\mathrm{HR}},\,\text{SDNN},\,\text{RMSSD}],
\]
where SDNN is the standard deviation of NN intervals, and RMSSD is the root mean square of successive differences.

\paragraph{HRV Derived from BVP Peaks (Dataset 3).}
When IBI was not available, HRV was computed by detecting systolic peaks in the BVP waveform. If peak timestamps are $\{p_j\}$:
\[
\mathrm{IBI}_j = p_{j+1}-p_j,
\]
\[
\text{SDNN}=\sigma_{\mathrm{IBI}},\qquad
\text{RMSSD}=\sqrt{\mathbb{E}[(\Delta\mathrm{IBI})^2]}.
\]

\paragraph{Tri-axial Accelerometer Features (ACC).}
Six motion features were computed from ACC-X, ACC-Y, and ACC-Z:
\[
f_{\mathrm{ACC}} = [\mu_x,\sigma_x,\mu_y,\sigma_y,\mu_z,\sigma_z].
\]

\paragraph{Blood Volume Pulse Features (Dataset 2).}
Two descriptors were included:
\[
f_{\mathrm{BVP}}=[\text{mean amplitude},\;\text{signal energy}].
\]

\paragraph{Final Multimodal Feature Vector.}
For subject $i$ and window $k$:
\[
\mathbf{f}_{i,k}=
[f_{\mathrm{EDA}},\, 
 f_{\mathrm{TEMP}},\,
 f_{\mathrm{HRV}},\,
 f_{\mathrm{ACC}},\,
 f_{\mathrm{BVP}},\,
 y_i ].
\]
Depending on dataset availability, this produces between 10--18 features per window.

\subsection{Feature Space Visualization and the Need for Nonlinear Modeling}

\begin{figure}[!htbp]
    \centering
    \includegraphics[width=0.95\linewidth]{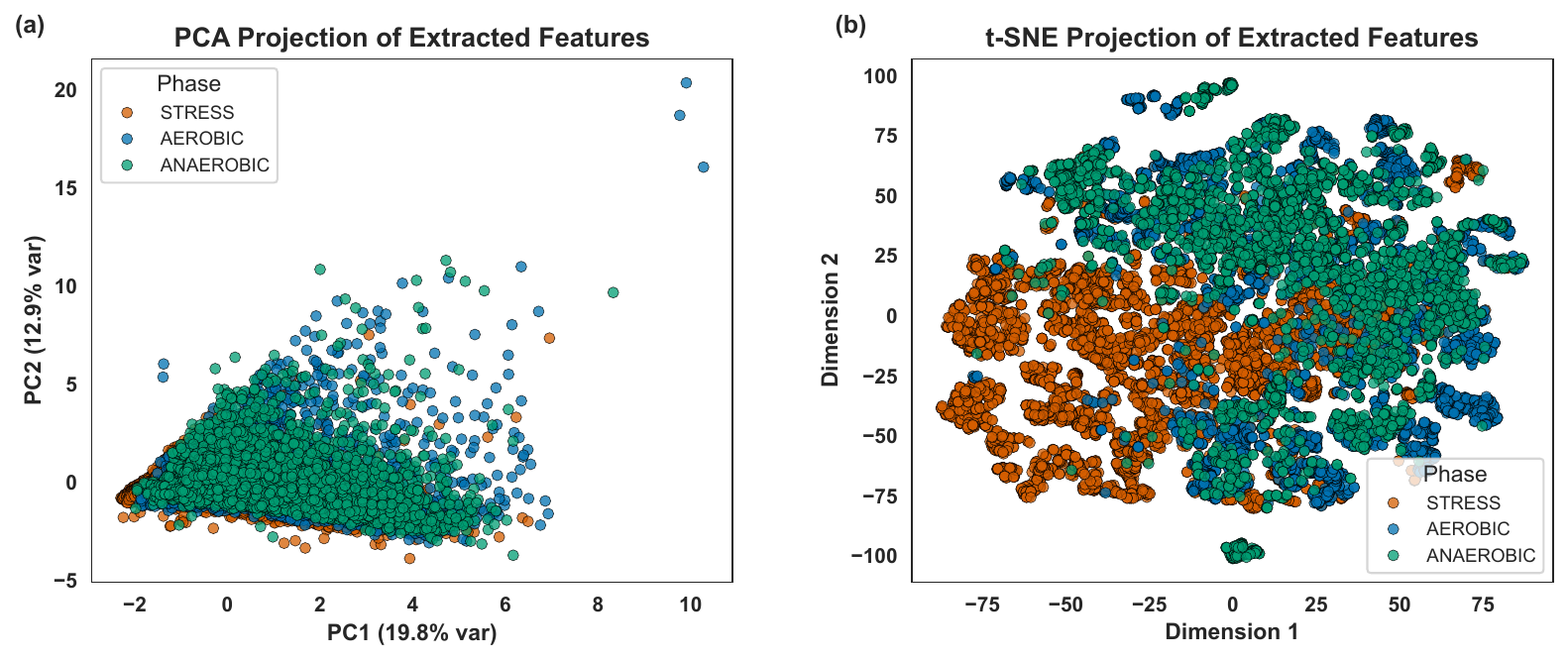}
    \caption{Pairwise feature-space projections for multimodal windows for Dataset~1.  
    Clear overlap between class clusters indicates that linear decision boundaries are insufficient to separate physiological states.}
    \label{fig:fsp1}
\end{figure}

\begin{figure}[!htbp]
    \centering
    \includegraphics[width=0.90\linewidth]{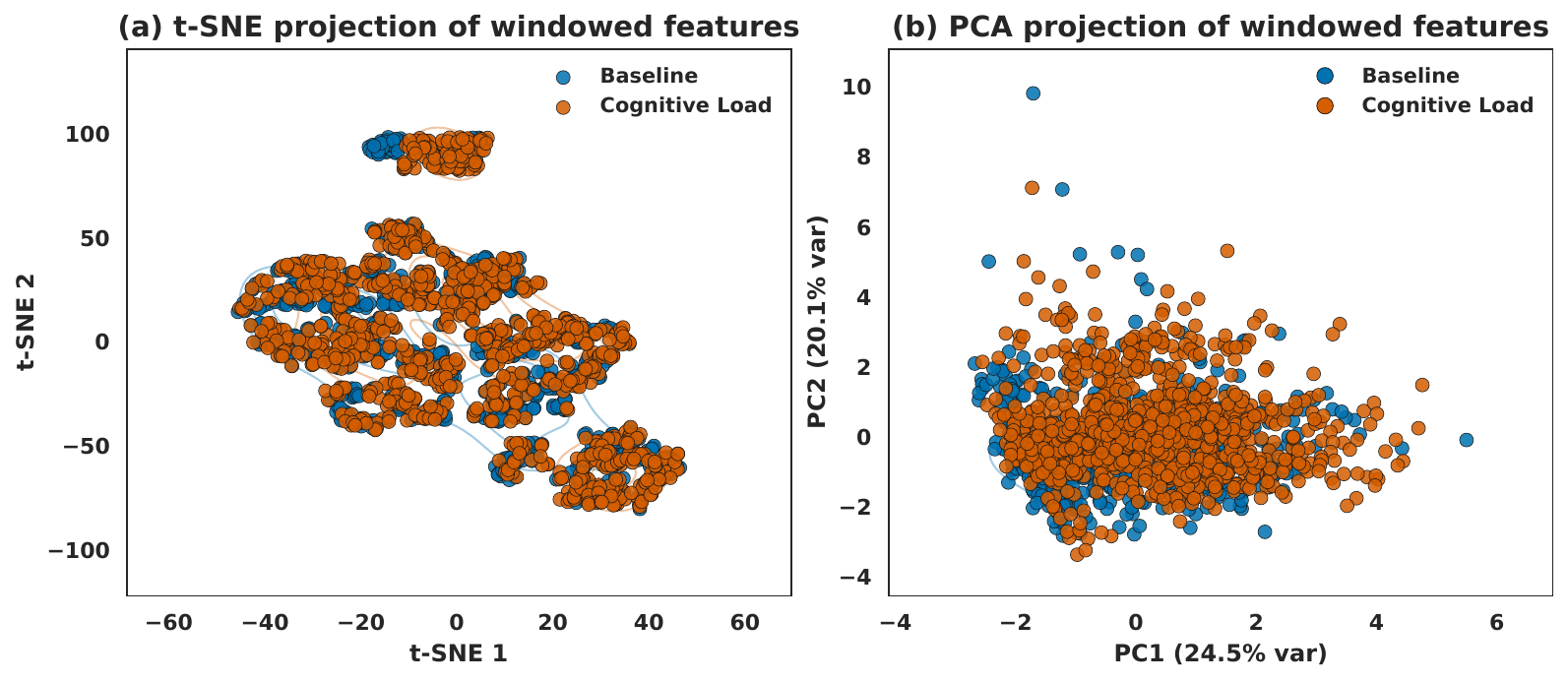}
    \caption{t-SNE and PCA embedding of the multimodal feature vectors for Dataset~2.  
    The highly curved, intertwined manifolds highlight nonlinear separability patterns across classes.}
    \label{fig:fsp2}
\end{figure}

\begin{figure}[!htbp]
    \centering
    \includegraphics[width=0.95\linewidth]{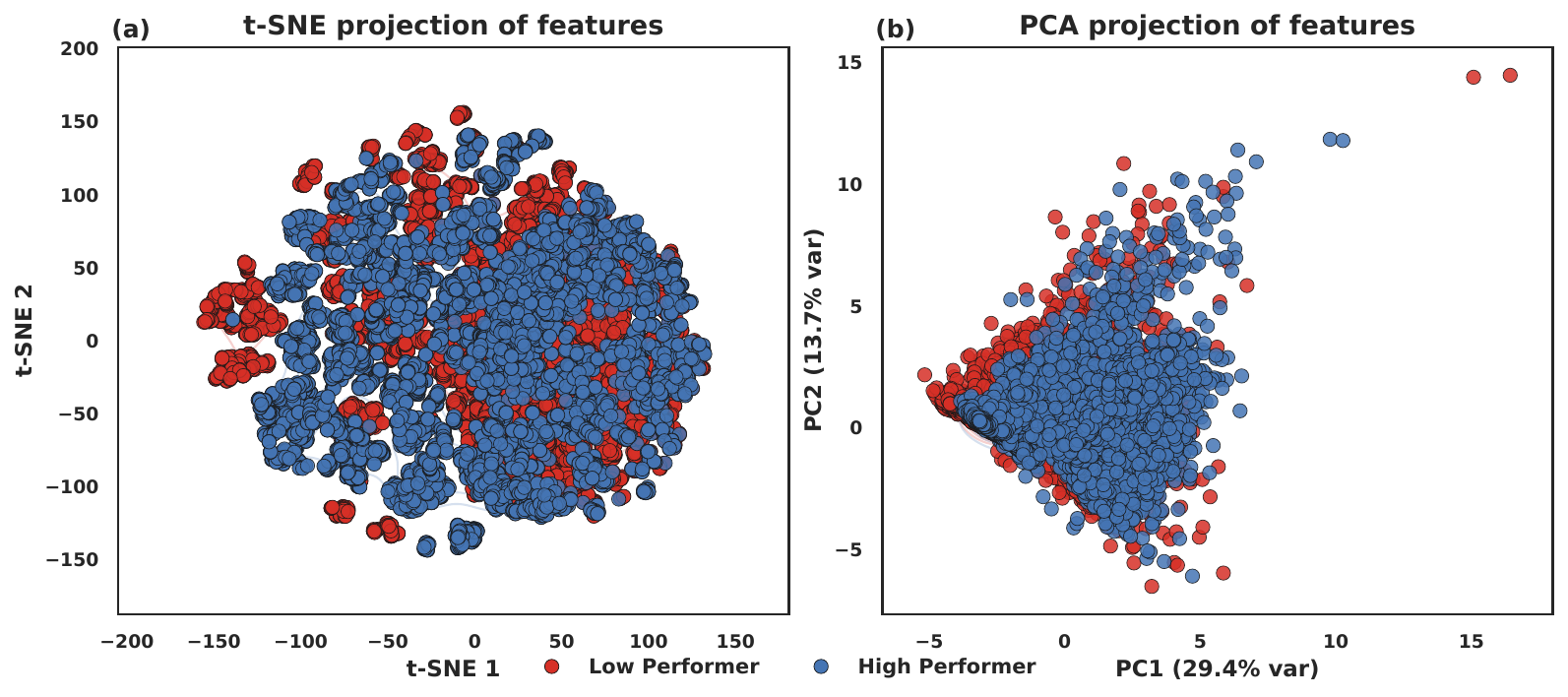}
    \caption{t-SNE and PCA projections overlaid with class labels for Dataset~3.   
    While PCA reveals substantial class overlap, t-SNE reveals local structure but still lacks linearly separable boundaries.}
    \label{fig:fsp3}
\end{figure}

To better understand the intrinsic structure of the extracted physiological features, three complementary feature space visualizations were generated using dimensionality reduction and pairwise projections. Figures~\ref{fig:fsp1}--\ref{fig:fsp3} illustrate how stress, cognitive load, and exercise-related states distribute within the multimodal feature space.

Across all three visualizations, a consistent pattern emerges: physiological responses to stress, cognitive load, and varying exercise intensities form complex, non-convex manifolds with substantial overlap in linear subspaces. In particular:
\begin{itemize}
    \item Pairwise feature projections (Fig.~\ref{fig:fsp1}) show intertwined spreads without clear linear margins.
    \item The t-SNE embedding (Fig.~\ref{fig:fsp2}) reveals curved manifolds, indicating class clusters that fold around each other.
    \item PCA (Fig.~\ref{fig:fsp3}) fails to provide meaningful separation, confirming that variance-based linear projections cannot disentangle autonomic and motion-derived features.
\end{itemize}
These observations collectively justify the use of nonlinear models such as SVM with RBF kernels, Random Forests, XGBoost, and LightGBM. Physiological patterns governing autonomic arousal, cardiovascular variability, and motion dynamics exhibit nonlinear interactions that cannot be captured by linear classifiers. This aligns with both the intrinsic biology of psychophysiological responses and the empirical performance gap observed in Section~\ref{sec:results}, where nonlinear models significantly outperform Logistic Regression across all datasets. Therefore, the feature space visualizations directly support the central claim of this study: nonlinear models are fundamentally necessary to capture the curved, high-dimensional manifolds underlying wearable physiological signals.

\subsection{Classification Models}

Following feature extraction, multiple nonlinear and linear machine learning models were evaluated to determine their ability to capture physiological dynamics underlying stress, cognitive load, exercise intensity, and performance level. Seven widely used classifiers were selected to represent distinct decision-making paradigms:
\begin{itemize}
    \item \textbf{Logistic Regression (LR):} A linear baseline model used to assess separability under linear decision boundaries.
    \item \textbf{Support Vector Machine (SVM, RBF kernel):} A nonlinear margin-based classifier capable of modeling complex physiological boundaries.
    \item \textbf{K-Nearest Neighbors (KNN):} A distance-based nonparametric classifier that adapts its decision boundaries to local geometric structure in the feature space.
    \item \textbf{Random Forest (RF):} An ensemble of decorrelated decision trees that captures nonlinear interactions and feature hierarchies.
    \item \textbf{Gradient Boosting Machine (GBM):} An iterative ensemble that builds shallow trees sequentially to correct prior errors, enabling strong nonlinear approximation.
    \item \textbf{XGBoost (XGB):} A scalable gradient-boosted tree model optimized for structured tabular data, effective in capturing subtle nonlinearities.
    \item \textbf{LightGBM (LGBM):} A leaf-wise boosted tree algorithm providing fast training and strong performance on high-dimensional feature spaces.
\end{itemize}
All hyperparameters were tuned using five-fold cross-validation within the training set. To avoid data leakage, windowed samples originating from the same subject were never split between training and validation folds.

\subsection{Evaluation Protocol}

To ensure robustness against inter-subject variability, two complementary subject-independent evaluation protocols were employed: (i) an 80/20 train--test split with complete subject separation to assess general performance, and (ii) LOSO cross-validation to quantify generalization to previously unseen individuals. Performance was evaluated using accuracy, precision, recall, F1-score, and AUC, with confusion matrices used to analyze class-specific error patterns. Statistical significance of performance differences was assessed using paired non-parametric tests with Bonferroni correction ($p < 0.05$)~\cite{sedgwick2012multiple}, confirming the reliability of observed trends.

Furthermore, an ablation study was conducted by retraining the best-performing nonlinear model after selectively removing individual physiological modalities. The results consistently demonstrated that multimodal fusion provides more discriminative power than unimodal configurations, highlighting the complementary nature of the physiological signals in recognizing cognitive and behavioral states.

\subsection{Interpretability Analysis}

Model interpretability was assessed using Shapley Additive Explanations (SHAP)~\cite{lundberg2017unifiedapproachinterpretingmodel}, a game-theoretic framework that attributes contributions of each feature to the model output. Global and class-wise SHAP analyses were performed. Averaging SHAP values across all test windows yielded a global ranking of the most influential physiological features, quantifying which modalities primarily drive model decision-making across all subjects. Class-wise SHAP beeswarm plots were generated to visualize how physiological features shift between cognitive load, stress, aerobic exercise, anaerobic exercise, and high/low performance. These patterns reveal modality-specific signatures, e.g., EDA slope for arousal, HRV metrics for cognitive load, and ACC variability for aerobic motion.

\begin{figure}[!htbp]
    \centering
    \includegraphics[width=0.95\linewidth]{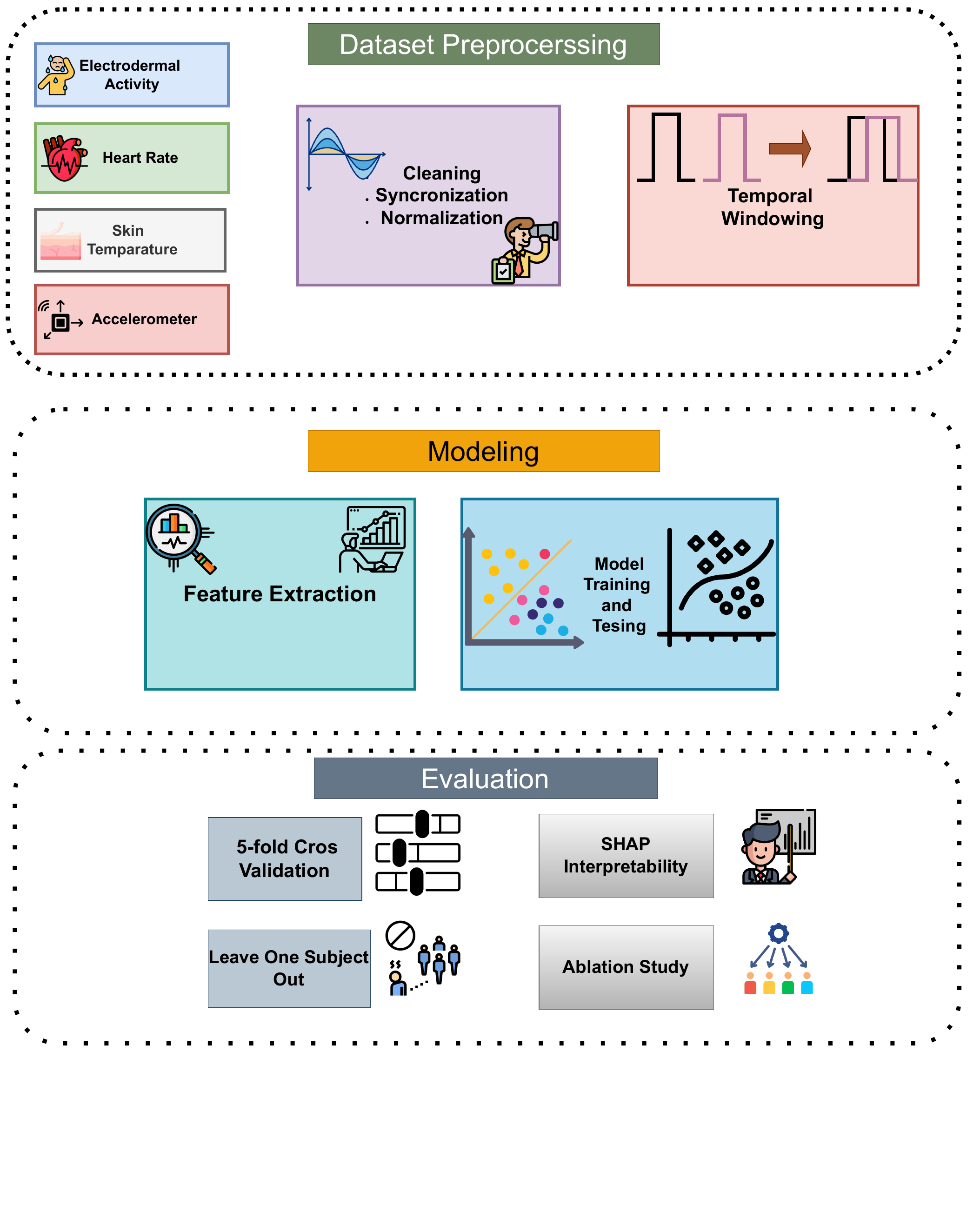}
    \caption{Overview of the proposed wearable physiological signal processing and modeling pipeline.}
    \label{fig:pipeline}
\end{figure}

Overall, SHAP-based interpretations provided physiologically meaningful insights, demonstrating that nonlinear models not only improved accuracy but also aligned with known autonomic, cardiovascular, and biomechanical responses underlying cognitive and physical states. As illustrated in Fig.~\ref{fig:pipeline}, the proposed framework encompasses the complete processing pipeline, including preprocessing, feature extraction, modeling, and evaluation.

\section{Results}
\label{sec:results}

This section reports the classification performance across the three wearable datasets. For all datasets, nonlinear models substantially outperformed the linear baseline (Logistic Regression), highlighting the inherent nonlinear structure of physiological responses.

\subsection{Dataset~1: Stress, Aerobic, and Anaerobic Classification}

Table~\ref{tab:results_dataset1} summarizes the performance of seven classifiers on the three-class stress--exercise dataset. Logistic Regression achieved only moderate performance (Accuracy = 0.636, Macro-F1 = 0.590), indicating that linear boundaries are insufficient for separating autonomic and motion-driven patterns. In contrast, nonlinear models exhibited a clear performance advantage: XGBoost achieved the highest accuracy (0.962) and Macro-F1 (0.958), followed closely by Random Forests (0.953), k-NN (0.895), and Gradient Boosting (0.884). The F1-scores for stress, aerobic, and anaerobic states remained consistently high for tree-based models.

\begin{table}[t]
\centering
\caption{Performance on the stress--exercise dataset (Dataset~1).}
\label{tab:results_dataset1}
\begin{tabular}{lcccccc}
\hline
\textbf{Model} & \textbf{Accuracy} & \textbf{MacroF1} & \textbf{WeightedF1} & \textbf{Stress F1} & \textbf{Aerobic F1} & \textbf{Anaerobic F1} \\
\hline
Logistic Regression & 0.636 & 0.590 & 0.621 & 0.776 & 0.493 & 0.502 \\
Random Forest       & 0.953 & 0.947 & 0.952 & 0.973 & 0.943 & 0.926 \\
SVM (RBF)           & 0.784 & 0.756 & 0.782 & 0.901 & 0.704 & 0.663 \\
kNN                 & 0.895 & 0.883 & 0.896 & 0.947 & 0.871 & 0.832 \\
Gradient Boosting   & 0.885 & 0.870 & 0.884 & 0.947 & 0.853 & 0.809 \\
\textbf{XGBoost}    & \textbf{0.962} & \textbf{0.958} & \textbf{0.962} & \textbf{0.979} & \textbf{0.956} & \textbf{0.940} \\
LightGBM            & 0.953 & 0.947 & 0.952 & 0.973 & 0.943 & 0.926 \\
\hline
\end{tabular}
\end{table}

These results demonstrate that physiological responses to stress and exercise intensities form nonlinear manifolds that are much better captured by ensemble-based classifiers.

\subsection{Dataset~2: Exam Performance Prediction}

Table~\ref{tab:results_dataset2} shows the results for predicting high versus low exam performance using multimodal physiological features extracted during the examination session. Again, nonlinear models substantially outperformed the linear baseline. LightGBM achieved the best overall performance (Accuracy = 0.977, F1 = 0.980, AUC = 0.998), with Random Forest and XGBoost showing similarly strong results.

\begin{table}[t]
\centering
\caption{Performance on the exam performance dataset (Dataset~2).}
\label{tab:results_dataset2}
\begin{tabular}{lccccc}
\hline
\textbf{Model} & \textbf{Accuracy} & \textbf{Precision} & \textbf{Recall} & \textbf{F1-score} & \textbf{ROC-AUC} \\
\hline
LightGBM & \textbf{0.977} & \textbf{0.982} & \textbf{0.978} & \textbf{0.980} & \textbf{0.998} \\
Random Forest & 0.973 & 0.978 & 0.975 & 0.977 & 0.997 \\
XGBoost & 0.973 & 0.979 & 0.973 & 0.976 & 0.997 \\
Gradient Boosting & 0.934 & 0.952 & 0.931 & 0.941 & 0.985 \\
kNN & 0.922 & 0.923 & 0.943 & 0.933 & 0.974 \\
SVM (RBF) & 0.906 & 0.899 & 0.941 & 0.920 & 0.961 \\
Logistic Regression & 0.661 & 0.660 & 0.835 & 0.737 & 0.671 \\
\hline
\end{tabular}
\end{table}

The sharp contrast between nonlinear and linear model performance confirms that exam stress and cognitive effort produce nonlinear, multimodal interaction patterns between ACC, EDA, TEMP, and BVP features.

\subsection{Dataset~3: Cognitive Load Classification}

Table~\ref{tab:results_dataset3} presents results for baseline versus cognitive load classification. LightGBM again yielded the highest accuracy (0.985) and F1-score (0.988), slightly outperforming Random Forest and XGBoost. The linear baseline achieved only 0.672 accuracy and 0.774 F1-score.

\begin{table}[t]
\centering
\caption{Performance on the cognitive load dataset (Dataset~3).}
\label{tab:results_dataset3}
\begin{tabular}{lccccc}
\hline
\textbf{Model} & \textbf{Accuracy} & \textbf{Precision} & \textbf{Recall} & \textbf{F1-score} & \textbf{ROC-AUC} \\
\hline
Logistic Regression & 0.6718 & 0.6798 & 0.8994 & 0.7739 & 0.6775 \\
SVM (RBF)           & 0.7241 & 0.7411 & 0.8590 & 0.7952 & 0.8009 \\
kNN                 & 0.7335 & 0.7641 & 0.8299 & 0.7953 & 0.7751 \\
Random Forest       & 0.9482 & 0.9527 & 0.9652 & 0.9589 & 0.9882 \\
Gradient Boosting   & 0.9747 & 0.9759 & 0.9840 & 0.9799 & 0.9948 \\
XGBoost             & 0.9800 & 0.9814 & 0.9868 & 0.9841 & 0.9971 \\
\textbf{LightGBM}   & \textbf{0.9853} & \textbf{0.9842} & \textbf{0.9925} & \textbf{0.9883} & \textbf{0.9989} \\
\hline
\end{tabular}
\end{table}

These results confirm that even in two-class settings where linear separability might be expected, nonlinear models yield clear advantages. Physiological responses to cognitive load induce subtle interactions across modalities that require nonlinear hypothesis spaces to be effectively captured.

Across all three datasets, a consistent pattern emerged:
\begin{itemize}
    \item Logistic Regression, the only linear model, performed substantially worse than all nonlinear alternatives.
    \item Tree-based models (Random Forest, XGBoost, LightGBM) achieved the highest performance in every dataset.
    \item The gap between linear and nonlinear models increased in multimodal settings involving HRV, BVP, and ACC.
\end{itemize}
These findings provide strong empirical support for the central claim of this work: nonlinear models are fundamentally necessary to capture the curved, multimodal physiological manifolds underlying stress, cognitive load, and performance-related responses.

\subsection{LOSO Validation}

\begin{figure}[!htbp]
    \centering
    \includegraphics[width=\linewidth]{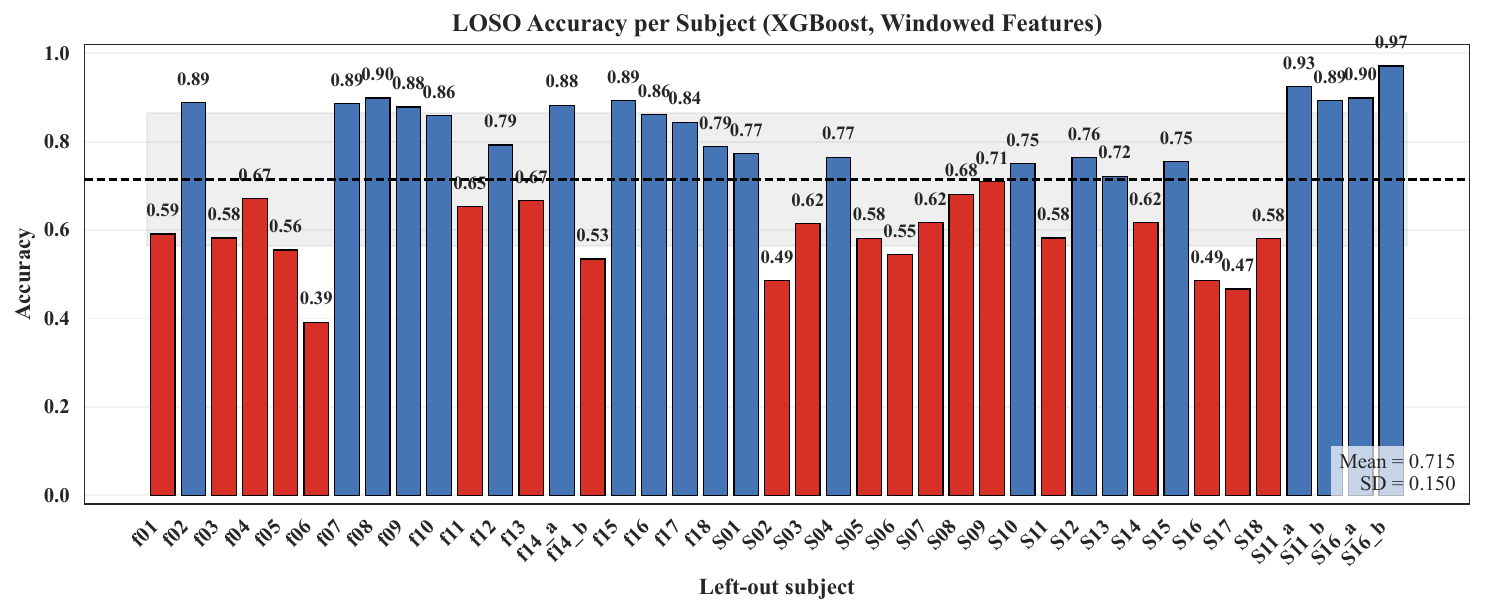}
    \caption{LOSO accuracy per subject for the stress--exercise dataset using XGBoost. The black dashed line denotes the mean accuracy (0.715). Substantial variability across subjects highlights the challenge of cross-person generalization in physiological modeling.}
    \label{fig:loso1}
\end{figure}

\begin{figure}[!htbp]
    \centering
    \includegraphics[width=0.95\linewidth]{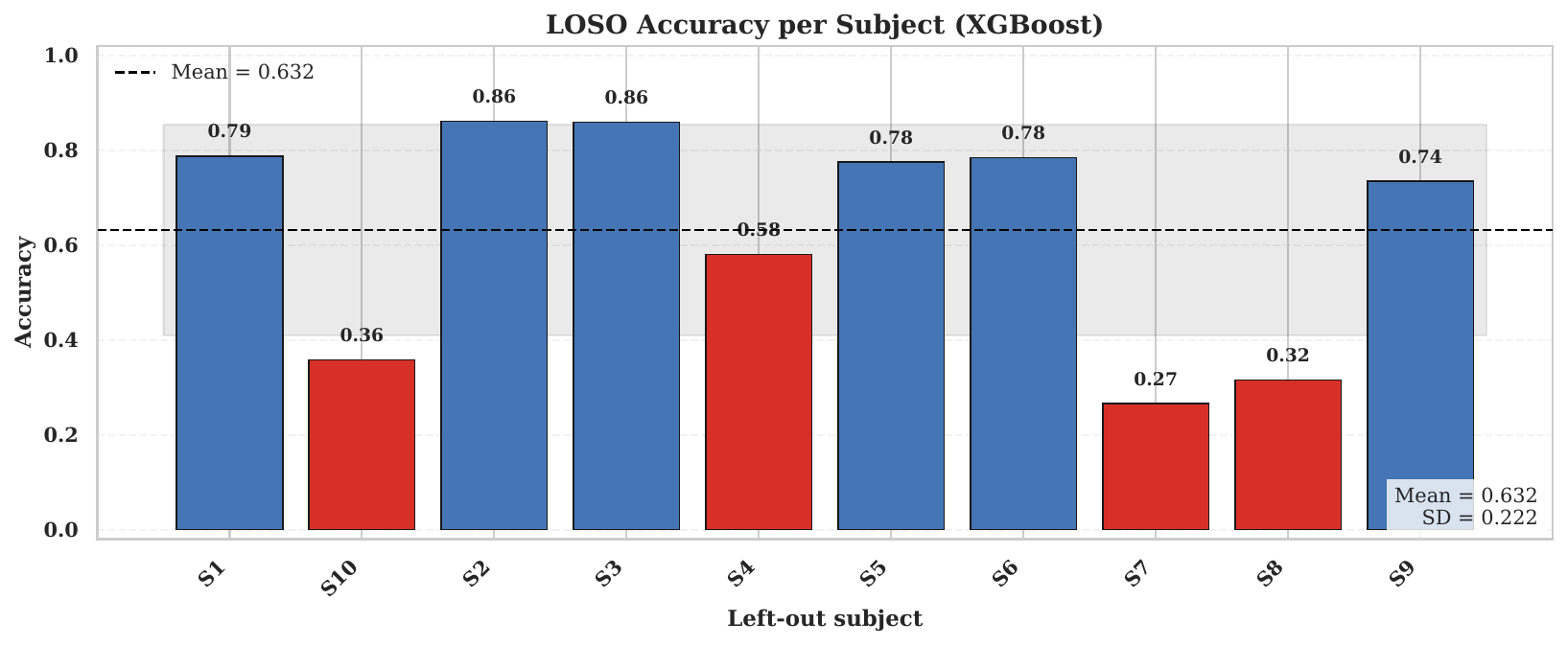}
    \caption{LOSO accuracy per subject for the exam performance dataset. The mean accuracy (0.632) is shown by the red dashed line. Some subjects exhibit high classification accuracy ($>$0.85), whereas others fall below 0.40, indicating large inter-individual differences in autonomic response patterns during evaluative stress.}
    \label{fig:loso2}
\end{figure}

\begin{figure}[!htbp]
    \centering
    \includegraphics[width=0.95\linewidth]{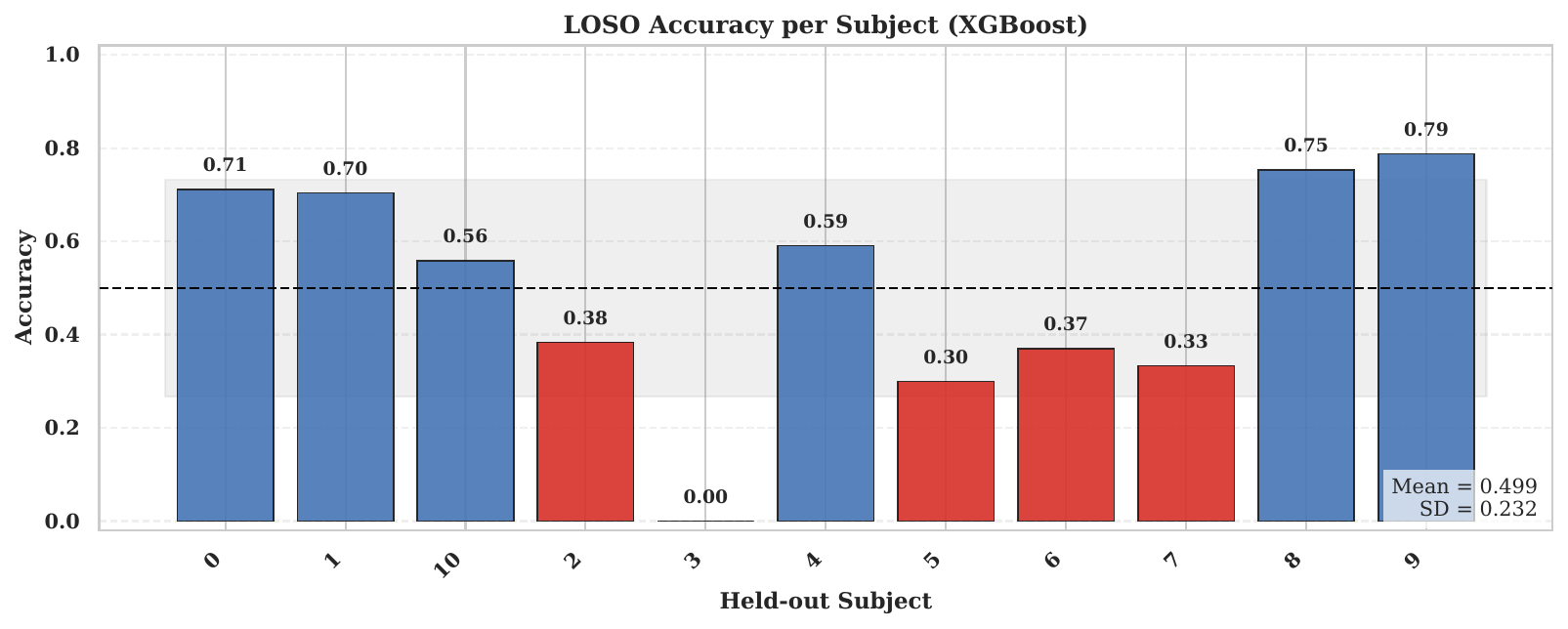}
    \caption{LOSO accuracy per subject for the cognitive load dataset. The mean accuracy (0.499) remains lower than the other datasets, reflecting the subtle, low-amplitude physiological changes associated with cognitive effort.}
    \label{fig:loso3}
\end{figure}

To evaluate generalization under strict subject independence, LOSO cross-validation was performed for all three datasets. In each fold, data from a single subject served as the test set, while the remaining subjects were used for training. Figures~\ref{fig:loso1}--\ref{fig:loso3} present the LOSO accuracy distributions across subjects for the stress--exercise dataset, exam performance dataset, and cognitive load dataset, respectively.

LOSO evaluation revealed substantial inter-subject variability across all datasets, with accuracy varying widely across participants. Cross-person generalization proved far more challenging than cross-session generalization, as mean LOSO accuracies (0.71, 0.632, 0.499) were considerably lower than subject-independent train--test results. Nevertheless, multimodal nonlinear models such as XGBoost maintained comparatively robust performance, indicating their ability to capture partially subject-invariant dynamics. Overall, these findings highlight strong subject-specific physiological signatures, motivating the need for personalization or domain adaptation strategies in future wearable systems.

\subsection{SHAP-Based Interpretability Analysis}

\begin{figure*}[!htbp]
    \centering
    \includegraphics[width=0.92\textwidth]{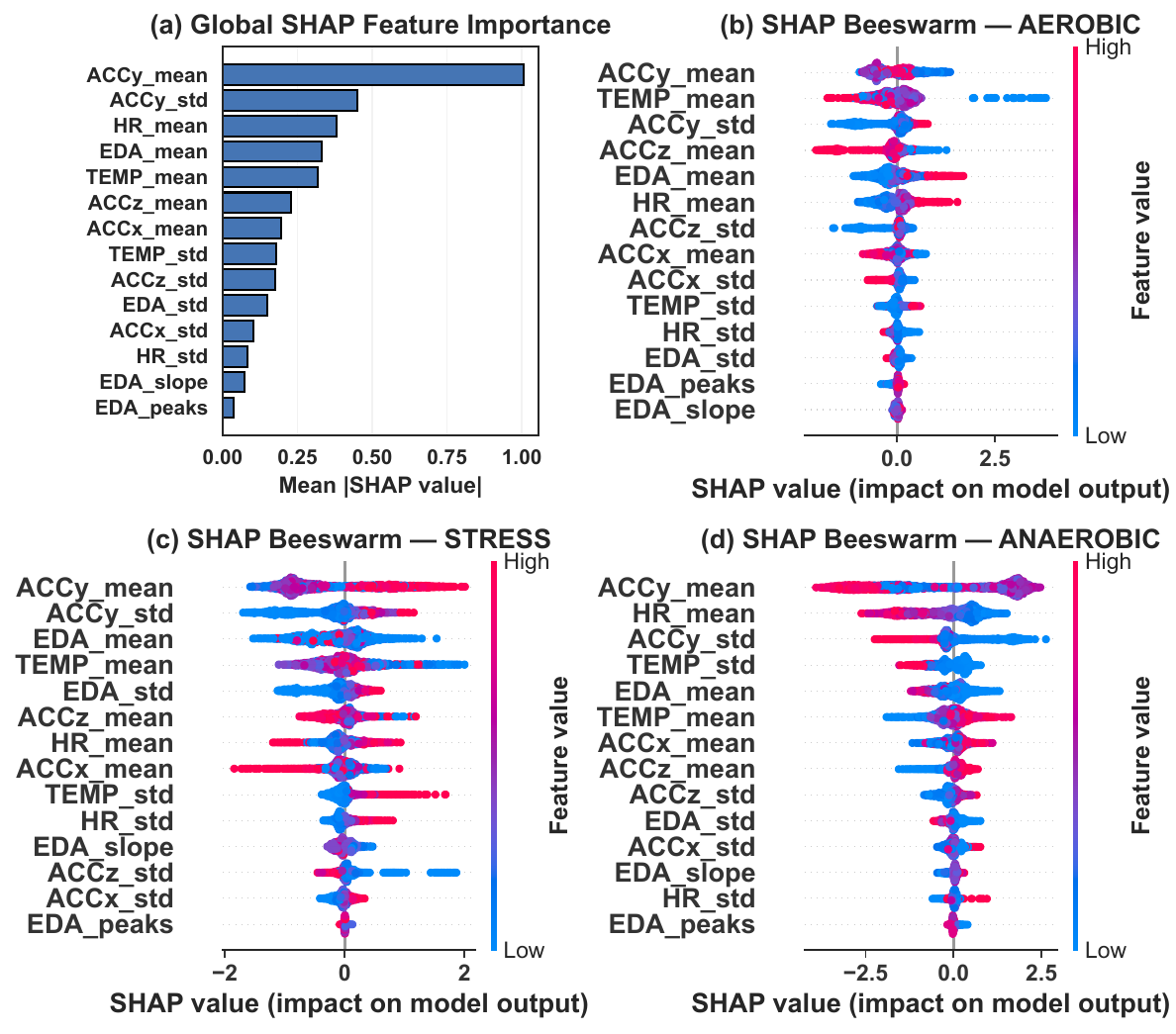}
    \caption{\textbf{SHAP interpretability for the exercise dataset.} Accelerometer-derived motion features dominate model decisions, and class-specific SHAP patterns reveal clear physiological signatures: aerobic activity is marked by moderate EDA and stable motion, stress by sympathetic activation (EDA and HR), and anaerobic effort by high motion variability and reduced HRV.}
    \label{fig:shap1}
\end{figure*}

\begin{figure*}[!htbp]
    \centering
    \includegraphics[width=0.98\textwidth]{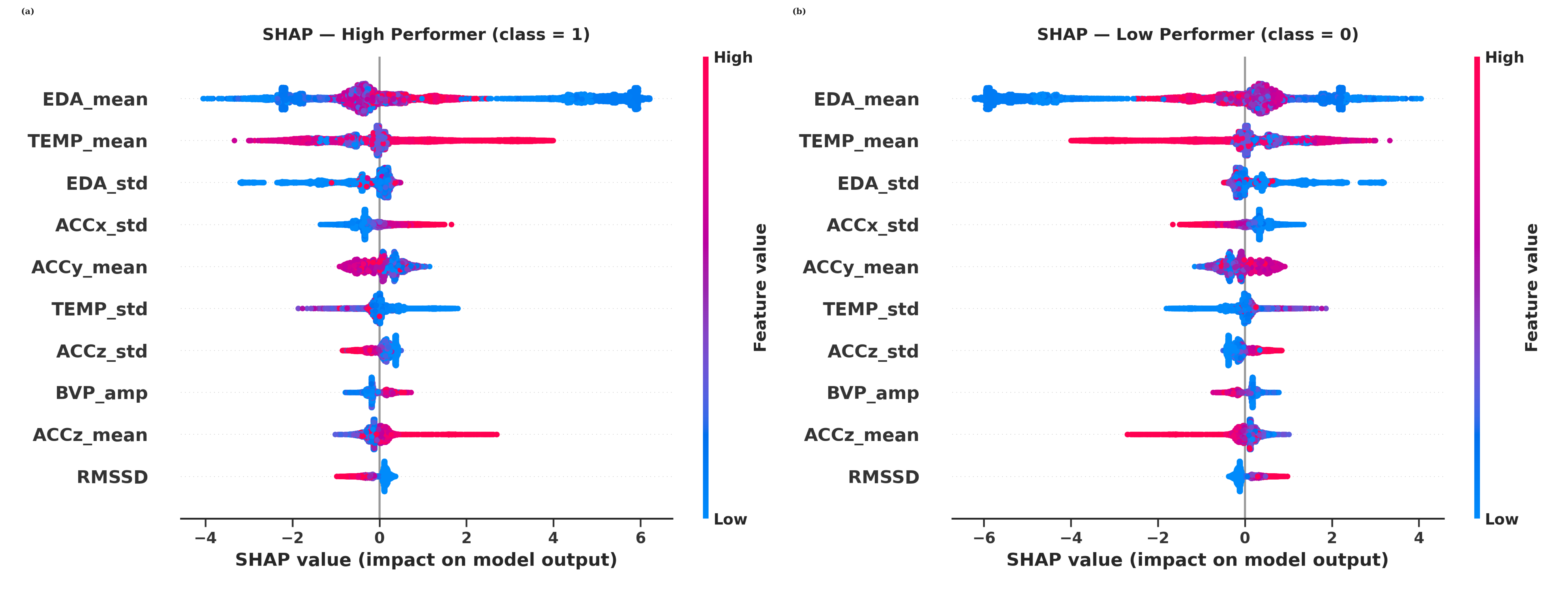}
    \caption{\textbf{SHAP analysis for cognitive load and performance prediction.} High-performing individuals show stable autonomic patterns driven by EDA\_mean, TEMP\_mean, and ACCy\_mean, while low performers exhibit increased variability in EDA, temperature, and motion, indicative of cognitive overload and fidgeting.}
    \label{fig:shap2}
\end{figure*}

\begin{figure*}[!htbp]
    \centering
    \includegraphics[width=0.92\textwidth]{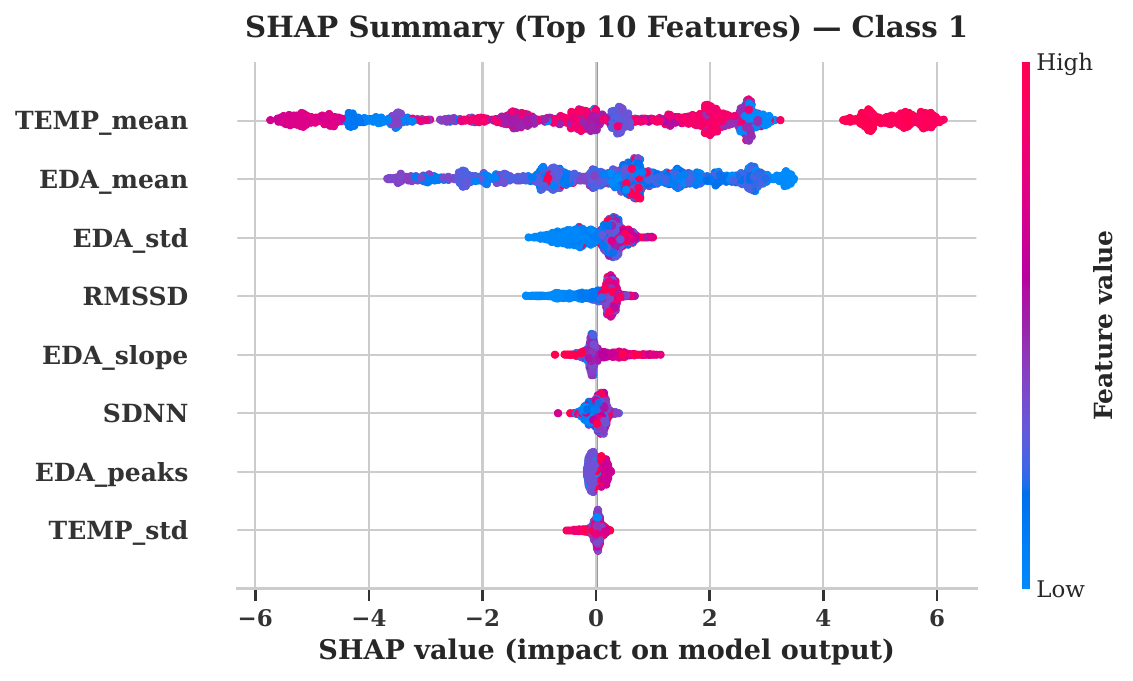}
    \caption{\textbf{SHAP interpretability for final performance classification.} Thermoregulatory (TEMP\_mean, TEMP\_std) and electrodermal (EDA\_mean, EDA\_std) features drive most predictions, while BVP variability reflects behavioral stability distinguishing baseline and cognitive performance.}
    \label{fig:shap3}
\end{figure*}

To examine how nonlinear models exploit multimodal physiological dynamics, SHAP analysis was performed across all three datasets. Figures~\ref{fig:shap1}--\ref{fig:shap3} provide global and class-specific explanations, revealing that the underlying decision boundaries are highly nonlinear and strongly dependent on feature interactions.

The global SHAP importance plot (Fig.~\ref{fig:shap1}, top-left) shows that accelerometer-derived features (ACC mean and variability) are the strongest predictors, likely reflecting micro-movement signatures linked to arousal, posture changes, and task engagement that linear models fail to capture. EDA metrics (EDA\_mean, EDA\_std) also exhibit strong influence due to sympathetic activation, while temperature and HRV features provide additional physiological context. Class-specific SHAP beeswarm plots (Fig.~\ref{fig:shap1}, right) reveal distinct response patterns: aerobic activity is characterized by structured motion and moderate increases in temperature and EDA; stress is driven by heightened EDA and elevated heart rate; and anaerobic exertion displays strong accelerometer variability and HRV suppression, indicative of explosive cardiovascular demand. These nonlinear, class-dependent feature signatures highlight the limitations of linear decision boundaries and explain the superior performance of nonlinear models in capturing multimodal physiological dynamics.

SHAP analyses from the cognitive performance and final achievement datasets (Figs.~\ref{fig:shap2}--\ref{fig:shap3}) reveal distinct autonomic and behavioral response strategies between high- and low-performing individuals. High performers show stable autonomic engagement characterized by consistently elevated EDA\_mean and TEMP\_mean with controlled movement (ACCy\_mean), whereas low performers exhibit larger fluctuations in EDA\_std, TEMP\_std, and ACCz\_std, indicative of stress-driven physiological instability and fidgeting under cognitive load. In Dataset~3, thermal and electrodermal synergy emerges as the dominant predictor: high EDA paired with reduced temperature is strongly associated with baseline rest tasks, while moderate activation alongside thermal stability favors cognitive load. Across datasets, SHAP value distributions demonstrate nonlinear, class-dependent, and interaction-driven feature effects, directly explaining the strong advantage of nonlinear models over linear approaches for multimodal physiological state recognition.

\subsection{Ablation Study and Statistical Analysis}

To evaluate the relative contribution of each physiological modality to classification performance, a comprehensive ablation study was conducted across the three datasets using a standardized XGBoost classifier and stratified cross-validation. For each dataset, the full multimodal feature set was compared (\textit{All}) against a series of ablated configurations in which specific modalities were removed (\textit{No\_EDA}, \textit{No\_TEMP}, \textit{No\_HR/HRV}, \textit{No\_ACC}, \textit{No\_BVP}) or isolated (\textit{Only\_EDA}, \textit{Only\_TEMP}, \textit{Only\_HR/HRV}, \textit{Only\_ACC}, \textit{Only\_BVP}). The goal of this analysis was twofold: (i) quantify the dependence of each dataset on individual sensor modalities and (ii) determine whether performance losses under ablation were statistically significant relative to the multimodal baseline.

Across all datasets, the full feature model consistently achieved the highest macro-F1 and accuracy values, underscoring the importance of multimodal fusion for robust physiological-state classification. Configurations excluding EDA (\textit{No\_EDA}) or temperature (\textit{No\_TEMP}) produced some of the largest performance reductions, reaffirming the relevance of these modalities in capturing sympathetic arousal and thermoregulatory responses. Conversely, removal of HR-derived features (\textit{No\_HR/HRV}) resulted in dataset-specific effects: the degradation was modest and not statistically significant for Dataset~3, whereas it was significant for Datasets~1 and~2. Ablating ACC features produced heterogeneous outcomes as well; performance dropped substantially in Dataset~1 but remained relatively high in Dataset~2, reflecting differences in locomotor activity between stress, aerobic exercise, and cognitive-load environments.

To assess statistical significance, per-fold performance was computed for differences between each ablation configuration and the full-feature baseline. Shapiro--Wilk tests~\cite{shapiro1965analysis} were applied to determine normality of the paired differences. Depending on normality outcomes, paired $t$-tests or Wilcoxon signed-rank tests were used. When applicable, false discovery rate (FDR) correction (Benjamini--Hochberg)~\cite{benjamini1995controlling} was applied to mitigate multiplicity effects. Cohen's~$d$~\cite{cohen2013statistical} was reported as an effect-size estimate for normally distributed comparisons.

\begin{table*}[t]
\centering
\caption{Ablation and statistical significance analysis for Dataset~1.
Macro F1 and accuracy are reported as mean~$\pm$~standard deviation across cross-validation folds.}
\label{tab:combined_d1}

\renewcommand{\arraystretch}{1.05}
\setlength{\tabcolsep}{3.8pt} % Reduce column separation
\small % Reduce font size (optional)

\begin{tabularx}{\textwidth}{%
l
c
c
c
c
c
c
c
c
c
}
\toprule
\textbf{Config} & 
\textbf{F1} & 
\textbf{Acc} &
\textbf{Shapiro $p$} &
\textbf{Test} &
\textbf{Stat} &
\textbf{$p_{\text{raw}}$} &
\textbf{$p_{\text{FDR}}$} &
\textbf{$d$} &
\textbf{Sig} \\
\midrule
All (14 features) & 0.908$\pm$0.007 & 0.919$\pm$0.006 & -- & -- & -- & -- & -- & -- & -- \\
No\_EDA & 0.880$\pm$0.010 & 0.894$\pm$0.008 & 0.537 & t-test & 15.99 & 8.95e--05 & 1.19e--04 & 7.15 & Yes \\
No\_TEMP & 0.863$\pm$0.007 & 0.879$\pm$0.007 & 0.073 & t-test & 34.32 & 4.30e--06 & 6.88e--06 & 15.35 & Yes \\
No\_HR & 0.899$\pm$0.008 & 0.910$\pm$0.007 & 0.577 & t-test & 10.70 & 4.32e--04 & 4.94e--04 & 4.79 & Yes \\
No\_ACC & 0.738$\pm$0.013 & 0.766$\pm$0.011 & 0.006 & Wilcoxon & 0.00 & 0.0625 & 0.0625 & 22.59 & No \\
Only\_EDA & 0.542$\pm$0.007 & 0.597$\pm$0.006 & 0.511 & t-test & 99.09 & 6.22e--08 & 2.49e--07 & 44.31 & Yes \\
Only\_TEMP & 0.498$\pm$0.016 & 0.561$\pm$0.012 & 0.413 & t-test & 62.37 & 3.96e--07 & 1.06e--06 & 27.89 & Yes \\
Only\_HR & 0.509$\pm$0.004 & 0.574$\pm$0.004 & 0.476 & t-test & 136.05 & 1.75e--08 & 1.40e--07 & 60.84 & Yes \\
Only\_ACC & 0.805$\pm$0.006 & 0.825$\pm$0.005 & 0.711 & t-test & 39.98 & 2.34e--06 & 4.68e--06 & 17.88 & Yes \\
\bottomrule
\end{tabularx}
\end{table*}

\begin{table*}[!htbp]
\centering
\caption{Combined ablation and statistical significance analysis for Dataset~2.
Macro F1 and accuracy are reported as mean~$\pm$~standard deviation across cross-validation folds.
Each ablation is statistically compared to the full feature model (All) using Shapiro--Wilk normality tests and paired \textit{t}-tests.
All configurations show statistically significant differences at $\alpha = 0.05$.}
\label{tab:combined_d2}
\begin{tabular}{lccccccc}
\hline
\textbf{Config} & \textbf{F1} & \textbf{Acc} & \textbf{Shapiro $p$} & \textbf{Test} & \textbf{$p$-value} & \textbf{Significant} \\
\hline
All (18 features) 
    & 0.967 $\pm$ 0.002 
    & 0.962 $\pm$ 0.002 
    & -- & -- & -- & -- \\
No\_EDA      
    & 0.940 $\pm$ 0.005 
    & 0.932 $\pm$ 0.005 
    & 0.172 & $t$-test & 7.14e--05 & Yes \\
No\_TEMP     
    & 0.940 $\pm$ 0.002 
    & 0.932 $\pm$ 0.002 
    & 0.096 & $t$-test & 1.44e--05 & Yes \\
No\_HR       
    & 0.964 $\pm$ 0.002 
    & 0.959 $\pm$ 0.003 
    & 0.901 & $t$-test & 2.12e--02 & Yes \\
No\_ACC      
    & 0.952 $\pm$ 0.004 
    & 0.946 $\pm$ 0.004 
    & 0.754 & $t$-test & 8.81e--04 & Yes \\
No\_BVP      
    & 0.963 $\pm$ 0.002 
    & 0.958 $\pm$ 0.003 
    & 0.229 & $t$-test & 2.87e--02 & Yes \\
Only\_EDA    
    & 0.843 $\pm$ 0.004 
    & 0.818 $\pm$ 0.005 
    & 0.298 & $t$-test & 1.43e--06 & Yes \\
Only\_TEMP   
    & 0.786 $\pm$ 0.008 
    & 0.747 $\pm$ 0.009 
    & 0.231 & $t$-test & 1.31e--06 & Yes \\
Only\_HR     
    & 0.720 $\pm$ 0.003 
    & 0.624 $\pm$ 0.004 
    & 0.384 & $t$-test & 2.03e--08 & Yes \\
Only\_ACC    
    & 0.894 $\pm$ 0.004 
    & 0.877 $\pm$ 0.004 
    & 0.211 & $t$-test & 1.92e--06 & Yes \\
Only\_BVP    
    & 0.704 $\pm$ 0.008 
    & 0.643 $\pm$ 0.009 
    & 0.781 & $t$-test & 3.70e--07 & Yes \\
\hline
\end{tabular}
\end{table*}

\begin{table*}[!htbp]
\centering
\caption{Ablation and statistical significance analysis for Dataset~3. 
Performance values are reported as mean~$\pm$~std across folds. 
Paired \textit{t}-tests with Shapiro--Wilk checks and FDR correction were applied.}
\label{tab:combined_d3}
\setlength{\tabcolsep}{4pt}
\begin{tabular}{lccccccc}
\hline
\textbf{Config} & \textbf{F1} & \textbf{Acc} & \textbf{Shapiro $p$} & \textbf{$t$-stat} & \textbf{$p$-FDR} & \textbf{$d$} & \textbf{Sig} \\
\hline
All (EDA+TEMP+HRV) 
    & 0.983$\pm$0.007 & 0.979$\pm$0.009 & -- & -- & -- & -- & -- \\
No\_EDA     
    & 0.905$\pm$0.012 & 0.878$\pm$0.016 & 0.664 & 10.92 & 0.00048 & 4.88 & Yes \\
No\_TEMP    
    & 0.886$\pm$0.010 & 0.854$\pm$0.012 & 0.068 & 14.51 & 0.00026 & 6.49 & Yes \\
No\_HRV     
    & 0.978$\pm$0.007 & 0.973$\pm$0.009 & 0.528 & 2.43  & 0.07192 & 1.09 & No \\
Only\_EDA   
    & 0.862$\pm$0.016 & 0.825$\pm$0.019 & 0.292 & 17.30 & 0.00020 & 7.74 & Yes \\
Only\_TEMP  
    & 0.886$\pm$0.015 & 0.855$\pm$0.018 & 0.509 & 11.81 & 0.00044 & 5.28 & Yes \\
Only\_HRV   
    & 0.736$\pm$0.012 & 0.629$\pm$0.012 & 0.484 & 29.23 & 0.00005 & 13.07 & Yes \\
\hline
\end{tabular}
\end{table*}

Taken together, the ablation and statistical analyses highlight several key findings: (i) multimodal feature fusion consistently enhances classification performance; (ii) EDA and temperature are among the most informative modalities across tasks; (iii) locomotor features (ACC) are indispensable for exercise-related classification but less critical for cognitive-load detection; and (iv) HRV is conditionally informative, with diminished benefit when sampling intervals or cognitive tasks limit autonomic variability.

\section{Discussion}

The results across all three datasets consistently demonstrate that nonlinear machine learning models are fundamentally better suited for wearable physiological state recognition than linear models. Tree-based ensembles (XGBoost, Random Forest, LightGBM) achieved high performance in cognitive load, exam stress, and exercise intensity tasks, with accuracies exceeding 0.95 in subject-independent evaluations and ROC-AUC values of nearly 0.99. In contrast, logistic regression struggled to exceed 0.70 AUC, particularly for stress and workload settings where multimodal interactions dominate. Feature-space visualizations further confirmed that feature manifolds are intertwined and non-convex, explaining why linear margins fail to separate states. LOSO validation highlighted strong inter-individual variability, reinforcing the challenge of generalizing across subjects but also showing that nonlinear methods retain moderate performance under unseen participants.

By applying a fully unified pipeline to heterogeneous datasets, this work reveals cross-domain invariants in physiological signaling: EDA dynamics consistently reflect sympathetic activation, ACC variability captures motion or fidgeting patterns, and temperature contributes to both exercise thermogenesis and stress-induced vasoconstriction. SHAP analysis provided mechanistic interpretability, exposing nonlinear interactions between modalities; for example, high EDA combined with temperature drops signaled poor exam performance, while anaerobic effort was marked by strong ACC variability and HRV suppression. These insights demonstrate that nonlinear models do not act as opaque black boxes, but instead align with known autonomic mechanisms.

The main novelty of this work lies in its cross-dataset benchmarking of nonlinear vs.\ linear models under a common framework, which is rarely explored in wearable sensing research~\cite{10.1007/978-981-16-0575-8_1,xu2023globem}. The findings converge toward a foundational conclusion: physiological responses to mental and physical demands are inherently nonlinear, and unified multimodal modeling provides stable biomarkers that generalize across contexts. Furthermore, the study emphasizes practicality by focusing on lightweight machine-learning techniques suitable for real-time, on-device inference, addressing a crucial gap between methodological sophistication and wearable deployability.

Despite strong results, several limitations remain. Physiological variability across users significantly degraded LOSO performance, indicating the need for personalization, domain adaptation, or calibration strategies~\cite{cosoli2022importance,yetton2019cognitive}. The cognitive-load dataset was small and produced marginal HRV effects due to its short task duration. Motion artifacts and environmental factors in the exam-stress dataset also restrict generalizability. Future work should incorporate larger, more diverse cohorts, deep temporal models that directly leverage waveform dynamics, and adaptive learning techniques that update to individual physiology over time. In addition, integrating contextual information and advanced artifact-removal pipelines may further enhance robustness for free-living deployment. Overall, this study establishes a promising foundation for reliable physiological state recognition and offers a unified benchmark to guide next-generation wearable health-monitoring systems.

\section{Conclusion}

This study provided a unified investigation of cognitive load, stress, and physical exercise recognition from wearable physiological signals, demonstrating that nonlinear models consistently outperform linear baselines across three publicly available Empatica~E4 datasets. The proposed pipeline, combining multimodal feature extraction, rigorous subject-independent evaluation, and SHAP-based interpretability, revealed that physiological state boundaries are fundamentally nonlinear and shaped by interactions among autonomic and motion-derived features. Multimodal fusion proved essential for robust classification, with EDA and temperature serving as the most informative modalities across tasks. Although strong performance was achieved in subject-separated evaluations, LOSO results emphasized persistent challenges in cross-person generalization due to subject-specific physiological signatures. These findings establish a methodological benchmark for lightweight, nonlinear models suitable for real-time wearable deployment, while motivating future research on personalization, domain adaptation, and validation in larger and more diverse free-living environments to fully realize reliable real-world physiological state recognition systems.

\section*{Acknowledgments}

The author acknowledges the use of digital writing tools (ChatGPT and Grammarly) solely for language refinement and expresses gratitude to the creators of the publicly available wearable physiology datasets and the developers of the open-source scientific computing libraries utilized in this study.

\section*{Funding}

This research did not receive any specific grant from funding agencies in the public, commercial, or not-for-profit sectors.

\section*{Author Contributions}

K.A.S.: Conceptualization, Methodology, Software, Experiments, Analysis, Writing -- Original Draft, Writing -- Review \& Editing.

\section*{Data Availability}

All datasets used in this study are publicly available from PhysioNet or the corresponding public repositories cited in the manuscript.

\section*{Supplementary Material}

Supplementary materials, code, and visualizations are available from the author upon reasonable request.

\bibliography{references}

\end{document}